\newif\ifpsfig
\psfigtrue
\ifpsfig
\documentstyle[psfig,preprint,tighten,floats,aps]{revtex}
\else\documentstyle[preprint,tighten,floats,aps]{revtex}\fi
\begin{document}
\title{Theoretical Model for the Semimetal \mbox{\boldmath Yb$_4$As$_3$}}
\author{Peter Fulde, Burkhard Schmidt, and Peter Thalmeier}
\address{Max-Planck-Institut f\"ur Physik komplexer Systeme, Dresden, Germany}
\date{April 7, 1995}
\preprint{MPI-PKS 95.001}
\maketitle
\begin{abstract}
We present a model which can explain semiquantitatively a number of the unusual
properties of \mbox{Yb$_4$As$_3$}. The structural phase transition at
$T_{\text{c}}\simeq300\,\mbox{K}$ is described by a band Jahn-Teller effect of
correlated electrons and is interpreted as a charge ordering of the Yb
ions. The low carrier concentration in the low-temperature phase follows from
the strong electron correlations of the 4f-holes on the Yb sites and can be
viewed as self-doping of charge-ordered chains. The observed heavy-fermion
behaviour is on a scale of $T^\ast\simeq50\,\mbox{K}$ and is due to spinon-like
excitations in the Yb$^{3+}$-chains. The appearance of a second low-energy
scale around 0.2\,K is due to the Fermi energy of the low-density carriers.
\end{abstract}

The rare-earth pnictide \mbox{Yb$_4$As$_3$} is a material with unusual physical
properties\cite{ref:bonville,ref:ochiai,%
ref:steglich,ref:suzuki,ref:suzuki:2,ref:ochiai:2}.
At $T_{\text{c}}\simeq300\,\mbox{K}$, a structural phase transition has been
observed which is accompanied by a charge ordering of the Yb ions and by
discontinuities of the resistivity and the Hall coefficient. The
high-temperature phase has the anti-\mbox{Th$_3$P$_4$} structure with space
group \mbox{I\=43d}, while the one of the trigonal low-temperature phase is
labelled \mbox{R3c}. M\"o\ss{}bauer spectroscopy data on \mbox{$^{170}$Yb} give
clear evidence that below 50\,K the system contains about 20\,\%
\mbox{Yb$^{3+}$} ions characterized by a \mbox{4f$^{13}$}-configuration and a
$J=7/2$ ground-state multiplett. This fraction is close to the expected value
of 25\,\% if perfect charge ordering takes place with the remaining 75\,\%
Yb-ions in a \mbox{Yb$^{2+}$}- or \mbox{4f$^{14}$}-configuration. No indication
of magnetic ordering is found down to 0.045\,K\cite{ref:bonville}. Instead, the
material shows typical heavy-fermion behaviour with a linear specific-heat
coefficient $\gamma\simeq200\,\mbox{mJ/mol\,K$^2$}$ and a correspondingly large
spin susceptibility $\chi_{\text{S}}$. The Sommerfeld-Wilson ratio is found to
be of order unity, clearly indicating Fermi-liquid behaviour. However, below
2\,K the susceptibility rises again, pointing to the presence of another
low-energy scale\cite{ref:bonville,ref:nakamura}. In agreement with the
Fermi-liquid behaviour the resistivity is found to be of the form
$\rho(T)=\rho_0+AT^2$ at low temperatures with a ratio $A/\gamma^\nu$
($\nu\simeq2$) similar to that of other heavy-fermion systems\cite{ref:ochiai}.

An unusual feature of \mbox{Yb$_4$As$_3$} is that the heavy-fermion properties
arise even though the density of charge carriers is extremely low. The
low-temperature Hall coefficient $R_{\text{H}}$ is positive, implying hole
conduction and has a value of
$(ecR_{\text{H}})^{-1}=7\cdot10^{18}\mbox{cm$^{-3}$}$. Thus the density of
charge carrier is of order 0.001/Yb-atom. Although \mbox{Yb$_4$As$_3$} is
often referred to as a ``low-carrier Kondo system'', it is very suggestive from
the above that the heavy-fermion like behaviour of the system must be of
different origin than in Kondo-lattice systems. (For reviews see, e.g.,
\cite{ref:suzuki:2,ref:kasuya}.) In fact, we are suggesting that it is
unrelated to the Kondo effect.

In the following we describe a model which can qualitatively (and
semiquantitatively) explain a number of the unusual properties of the material,
in particular the low carrier-density. It differs from the one suggested in
\cite{ref:ochiai} which is based on a slightly depleted pnictogen band crossing
a narrow Yb 4f band of width $\sim50\,\mbox{K}$. (The latter is assumed to have
one hole per eight Yb ions.) Our model is based on a band Jahn-Teller effect of
correlated electrons ({\sc cbjt}) and rests on the following basic assumptions:
\begin{itemize}
\item The fifty-six Yb 4f-bands have an overall width of $\sim0.2\,\mbox{eV}$
which arises from effective f-f hopping via hybridization with pnictide valence
states\cite{ref:takegahara}. In \mbox{Yb$_4$As$_3$} all Yb-atoms are located on
four families of chains oriented along the space diagonals of the cubic unit
cell\cite{ref:okeeffe}. To reduce the complexity of the f-bands we make the
assumption that they
can be described by four degenerate one-dimensional bands associated with the
chains. Such a model is rather similar to the Labb\'e-Friedel model for 3d
states in \mbox{A\,15} compounds\cite{ref:labbe-friedel} where one has chains
parallel to the cubic axes. Although such models may not be literally true due
to interchain coupling, they describe the important aspect that a strain
coupling to the one-dimensional band states may easily lead to distortions of
the cubic structure and simultaneous repopulation among 4f band states of
\mbox{Yb$_4$As$_3$}.
\item There is a strong deformation potential coupling typical for mixed-valent
systems which removes this degeneracy by a trigonal {\sc cbjt}-distortion. When
the Jahn-Teller transition takes place, the crystal shrinks in
$\langle111\rangle$-direction, lifting the four-fold degeneracy. The four
equivalent chains are subdivided into one along the
$\langle111\rangle$-direction, for simplicity referred to as {\em short
chain}, and the three remaining chains referred to as {\em long chains}. Since
the \mbox{Yb$^{3+}$} ions have a smaller radius than the \mbox{Yb$^{2+}$} ones,
it is natural to think of the {\sc cbjt} transition being related to the
ordering of \mbox{Yb$^{3+}$} ions along the space diagonal $\langle111\rangle$.
\end{itemize}

We describe the {\sc cbjt} transition by an effective Hamiltonian of the form
\begin{mathletters} \label{eq:H}
\begin{equation} \label{eq:Ha}
	H=-t\sum_{\mu=1}^{\mu_{\text{f}}}\sum_{\langle ij\rangle\sigma}
		f_{i\mu\sigma}^\dagger f_{j\mu\sigma}
		+\epsilon_\Gamma\sum_{\mu=1}^{\mu_{\text{f}}}\Delta_\mu
		f_{i\mu\sigma}^\dagger f_{i\mu\sigma}
		+N_{\text{L}}\mu_{\text{f}}c_0\epsilon_\Gamma^2,
\end{equation}
where $\mu$ labels the summation over the $\mu_{\text{f}}=4$ different chains,
and $\langle ij\rangle$ denotes a summation over nearest-neighbour sites along
one chain. The $f_{i\mu\sigma}^\dagger$ create f-holes with spin $\sigma$ at
site $i$ of chain $\mu$. We choose a bandwidth $4t=0.2\,\mbox{eV}$ as obtained
from {\sc lda} calculations\cite{ref:takegahara} and, for simplicity, an
effective spin degeneracy of $2S+1=2$. The second term in (\ref{eq:Ha})
describes the volume-conserving coupling of the trigonal strain
$\epsilon_\Gamma$ ($\Gamma\equiv\Gamma_5$) to the f-bands characterized by a
deformation potential
\begin{equation}
	\Delta_\mu=\Delta\left(\delta_{\mu1}-(1-\delta_{\mu1})
		\frac{1}{\mu_{\text{f}}-1}\right).
\end{equation}
\end{mathletters}
We assume that changes of the bandwidths due to the distortion are small and
can be neglected. The third term in (\ref{eq:Ha}) is the elastic energy
associated with the distortion. Here $N_{\text{L}}$ is the number of sites in a
chain and $c_0$ is the background elastic constant for one chain for which we
choose $c_0/\Omega=10^{11}\,\mbox{erg/cm$^3$}$, where $\Omega$ is the volume of
a unit cell. (The lattice constant is $a=8.789\,\mbox{\AA}$.) Note that the
bulk elastic constant $c_\Gamma^0=4c_0$. An elastic constant of this size is
common for rare-earth systems, but for \mbox{Yb$_4$As$_3$} it has not yet been
measured. The Hamiltonian (\ref{eq:H}) does not yet contain the strong Coulomb
interactions of holes at an Yb site. Therefore it is reasonable only above
$T_{\text{c}}$ where the number of holes per Yb site is 1/4 and somewhat below
$T_{\text{c}}$. At low temperatures the Coulomb interactions and the strong
correlations which they imply are crucial (see below). However, for modeling
the {\sc cbjt} phase transition the Hamiltonian (\ref{eq:H}) is sufficient. The
condition for a phase transition to occur is $\Delta^2/(4tc_0)>3$. We choose
$\Delta=5\,\mbox{eV}$, which corresponds to a reasonable Gr\"uneisen parameter
$\Omega\equiv\Delta/(4t)=25$. From this, a transition temperature of
$T_{\text{c}}\simeq250\,\mbox{K}$ is obtained which is close to the observed
$T_{\text{c}}^{\text{exp}}\simeq300\,\mbox{K}$.

\begin{figure}
\ifpsfig\centerline{\psfig{figure=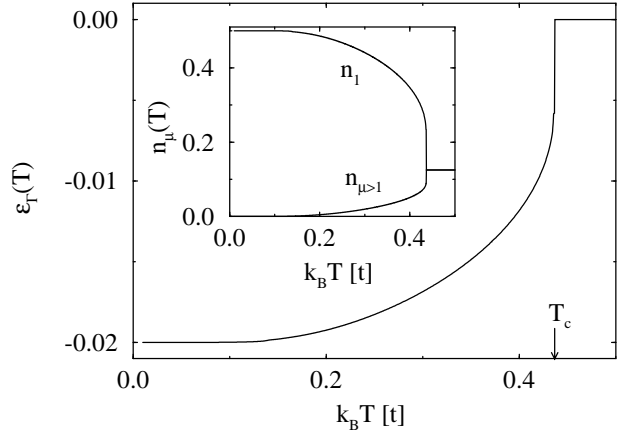,width=8cm,angle=-90}}
\else\centerline{\hbox{\fbox{\vbox to 5.87cm{\vss\hbox to 8cm{\hss}\vss}}}}\fi
\vspace{0.5in}
\caption{Temperature dependence of the strain order parameter
$\epsilon_\Gamma(T)$. Inset: Occupation numbers $n_\mu(T)$ of the four
f-bands.}
\label{fig:bs1}
\end{figure}

Below the phase transition, one obtains a strain order parameter
$\epsilon_\Gamma(T)$ whose temperature variation together with the band
occupation $n_\mu(T)$ is shown in Fig.~\ref{fig:bs1}.
At $T_{\text{c}}$, the four degenerate bands split with an associated shift of
holes from the upper threefold degenerate bands into the lower fourth band. The
energy difference between the respective band centers is
$\frac{4}{3}|\epsilon_\Gamma\Delta|$. The equilibrium strain at low temperature
is $\epsilon_\Gamma=-\Delta/(8c_0)\simeq-0.02$. The increase in the hole
occupation number of the lower band shows a similar behaviour as
$|\epsilon_\Gamma(T)|$ until at low $T$ the upper bands are empty (of holes)
while the lower band is becoming half-filled.

As the lower band is approaching half-filling, the strong on-site f electron
correlations become more and more important. Together with the coupling between
short and long chains, this has important consequences as we will
demonstrate. We incorporate the strong correlations by an on-site Hubbard
interaction $U$ which can be eliminated by going over to a t-J Hamiltonian of
the form
\begin{equation} \label{eq:H_1}
	H^{(1)}=-t\sum_{\langle ij\rangle\sigma}{\cal P}f_{i1\sigma}^\dagger
		f_{j1\sigma}{\cal P}
	+J\sum_{\langle ij\rangle}\left({\bf S}_i^{(1)}\cdot{\bf S}_j^{(1)}
		-\frac{1}{4}n_i^{(1)}n_j^{(1)}\right),
%	+\Delta\epsilon_\Gamma\sum_{i\sigma}n_{i\sigma}^{(1)}
%	+N_{\text{L}}c_0\epsilon_\Gamma^2
\end{equation}
where $J=4t^2/U$ and ${\bf S}_i^{(\mu)}=\sum_{\alpha\beta}
f_{i\mu\alpha}^\dagger\mbox{\boldmath$\sigma$}_{\alpha\beta}f_{i\mu\beta}$,
$n_i^{(\mu)}=\sum_\sigma n_{i\sigma}^{(\mu)}$. With $U=10\,\mbox{eV}$, we
obtain $J=1\cdot10^{-3}\,\mbox{eV}$. The projector $\cal P$ serves for the
projection onto singly occupied sites.  We make use of the so-called
slave-boson approximation\cite{ref:read,ref:lee}, representing the projector
$\cal P$ by auxiliary bosons $b_j$.  The projection back onto the physical
states of the now enlarged Hilbert space is done by means of Lagrange
multipliers $i\lambda_j$ for each site $j$.  We introduce two
Hubbard-Stratonovi\v{c} fields $\chi_{ij}=\langle \sum_\sigma
f_{i1\sigma}^\dagger f_{j1\sigma}\rangle$ and $r_j= \langle b_j\rangle$ to
decouple the Hamiltonian (\ref{eq:H_1}), assuming that we work in a range of
parameters where we can ignore Cooper pairing by setting $\langle
f_{i1\uparrow}^\dagger f_{j1\downarrow}^\dagger-f_{i1\downarrow}^\dagger
f_{j1\uparrow}^\dagger\rangle=0$.  ($\langle{\bf S}_j^{(1)}\rangle=0$ as well.)
Performing the uniform-saddle point approximation, $\chi_{ij}=\chi$, $r_j=r$,
$i\lambda_j=\lambda$, the Hamiltonian becomes after Fourier transformation
\begin{mathletters} \label{eq:H_MF}
\begin{eqnarray}
	H_{\text{MF}}^{(1)}&=&\sum_{k\sigma}\xi_kf_{k1\sigma}^\dagger
		f_{k1\sigma}
	+N_{\text{L}}\left(\frac{3}{4}J\chi^2-\lambda\left(1-r^2\right)
		-\frac{1}{2}J\left(1-r^2\right)^2
%		+c_0\epsilon_\Gamma^2
		\right),
\\
	\xi_k&=&-2\left(r^2t+\frac{3}{4}J\chi\right)\cos k+\lambda
	+\frac{3}{4}J
%	+\Delta\epsilon_\Gamma
\end{eqnarray}
\end{mathletters}
The variational equations for the mean-field parameters lead to $r^2=\delta$
(where $\delta$ is the deviation from half-filling of chain 1),
$\lambda=2t\chi$, and
$\chi(T=0)=\frac{2}{\pi}\sin\left(\frac{\pi}{2}(1-\delta)\right)$.  We observe
that for low temperatures, the thermodynamics is essentially governed by a
renormalization of the effective mass of the lower band,
\begin{equation} \label{eq:m_ast}
	\frac{m^\ast}{m_{\text{b}}}=\frac{t}{\delta t+\frac{3}{4}\chi J}
		\simeq1.0\cdot10^2,
\end{equation}
where $m_{\text{b}}$ is the bare band
mass. This value for $m^\ast$ corresponds to a spinon bandwidth of 25\,K, which
is close to the experimental $T^\ast$.

This brings us to an important point, namely the determination of the quantity
$\delta$ in the limit of $T\to0$. Without interchain coupling, the upper three
bands are empty (of holes) in that limit, while the lower band is
half-filled. The system should be a Mott-Hubbard insulator in that case. This
changes when the interchain coupling is taken into account. In order to
appreciate this point we neglect first the correlations. We introduce
transversal hopping matrix elements $t_\bot'\epsilon_\Gamma$ induced by the
distortion between the short chain and the other three chains.  These are again
effective matrix elements, because the Yb-Yb hopping is viewed as taking place
via intermediate pnictogen states. We have scaled $t_\bot'$ with the strain to
be able to restore the cubic symmetry at $T\geq T_{\text{c}}$. Then we have to
find new single-particle eigenstates $\phi_{k\lambda\sigma}^\dagger=\sum_\mu
e_{\lambda\mu}^\ast(k)f_{k\mu\sigma}^\dagger$ and the corresponding
eigenvalues. The latter result is an additional repulsion of the lower and
upper bands by an amount of $9t_\bot^{\prime2}/(4\Delta)$. The lower band is
still half-filled, but when the strong correlations are taken into account,
this does no longer imply a Mott-Hubbard insulating behaviour. Instead, the
hole count on the short chain is reduced to
\begin{equation} \label{eq:occ-num}
	n_{\mu=1}=\frac{1}{N_{\text{L}}}\sum_{k\sigma}
		\left|e_{11}(k)\right|^2\left\langle\phi_{k1\sigma}^\dagger
		\phi_{k1\sigma}\right\rangle\simeq1-
		\frac{27}{16}\left(\frac{t_\bot'}{\Delta}\right)^2,
\end{equation}
since the single-particle eigenstates $\phi_{k\lambda\sigma}$ contain
contributions $\propto t_\bot'/\Delta$ from the long chains. The Hubbard-$U$
has an effect only in the short chains where the hole concentration is nearly
1, but not in the long chains where the hole density is very small. Therefore
the system acts like being self-doped with a doping parameter
$\delta=\frac{27}{16}\left(\frac{t_\bot'}{\Delta}\right)^2$. Taking
$\delta\simeq0.6\cdot10^{-3}$ from experiment, we have
$t_\bot'\simeq0.1\,\mbox{eV}$. Note that the self-doping found here differs
from the one, e.g., in \mbox{YBa$_2$Cu$_3$O$_7$} where holes are transferred
from the \mbox{Cu$^{3+}$} ions in the chains to the \mbox{Cu$^{2+}$} ions in
the planes because several bands are crossing the Fermi level.

With the picture outlined above one can understand, at least qualitatively, a
number of properties of \mbox{Yb$_4$As$_3$}. The large $\gamma$-coefficient
results from the (spinon-like) excitations within the short chains. We obtain
it formally from the large mass enhancement (\ref{eq:m_ast}). The same holds
true for the spin susceptibility. At this point it should be noted that our
model contains a second temperature scale, which is given by the Fermi energy
$\varepsilon_{\text{F}}^{\text{h}}$ of the holes in the long chains. For
$\delta\simeq0.6\cdot10^{-3}/\mbox{Yb-atom}$ we find
$\varepsilon_{\text{F}}^{\text{h}}\simeq0.2\,\mbox{K}$. This might explain the
low-temperature behaviour of $\chi_{\text{S}}(T)$, because the holes in the
long chains give rise to a Curie-like susceptibility except at
$k_{\text{B}}T\ll\varepsilon_{\text{F}}^{\text{h}}$ when the Fermi gas becomes
degenerate.

As far as transport is concerned, we expect that the same heavy quasiparticles
responsible for the large $\gamma$-coefficient are also governing the low-$T$
behaviour of the resistivity. The resistivity of the short chains is presumably
higher than the one of the holes in the long chains. In a crystal with
different domains the one of the short chains will therefore make the largest
contribution to the measured resistivity. This might explain why the prefactor
$A$ of the $T^2$-term leads to a value for the ratio $A/\gamma^2$ like in other
heavy-fermion systems\cite{ref:kadowaki}. As far as $\rho(T)$ for $T\lesssim
T_{\text{c}}$ is concerned, it must increase with decreasing temperature
because of the depletion of holes in the upper bands. Thus $\rho(T)$ must have
a maximum in order to match on to the $\rho_0+AT^2$ behaviour at low
$T$. Concerning the Hall coefficient, we expect that in the presence of
different channels with different conductivities $\sigma_\mu$ the one with the
largest $\sigma$ makes the biggest contribution\cite{ref:ziman}. This would
imply that the holes in the long chains are most effective here. Since the hole
density increases with rising $T$, the Hall coefficient decreases
accordingly. The initial increase for small temperatures is possibly an effect
beyond the simple one-band model expression for the Hall coefficient. It is
worth noticing that in the high-temperature phase the experimental Hall
coefficient is $(ecR_{\text{H}})^{-1}=1\cdot10^{21}\,\mbox{cm$^{-3}$}$. With a
volume $\Omega=6.8\cdot10^{-22}\mbox{cm$^3$}$ per unit cell, this would
correspond to 0.25 holes per formula unit \mbox{Yb$_4$As$_3$} provided we have
a one-band system. With four independent bands the application of a generalized
theory for the Hall coefficient\cite{ref:ziman} yields one hole per formula
unit, as expected from a chemical approach using valence electron counting.

In conclusion we may state that the model presented here for
\mbox{Yb$_4$As$_3$} explains semiquantitatively the most important experimental
findings in a natural way. Among them are the low carrier concentration in the
low-temperature phase, the heavy-fermion behaviour with an effective mass ratio
$m^\ast/m\simeq100$ as well as the appearance of a second low-energy scale of
order 0.2\,K.

\acknowledgements
We thank Dr.~G. Zwicknagl for discussions on low-carrier systems and
Prof.~Dr.~H.~G. v.~Schnering and Prof.~Dr.~A. Simon for elucidating comments
on the crystal structure of \mbox{Yb$_4$As$_3$}.

\end{document}